\documentclass{article}
\usepackage[utf8]{inputenc}
\usepackage{amsmath,amsfonts,amssymb}
\usepackage{geometry}
\usepackage{natbib}       % Enhanced bibliography management
\usepackage{hyperref}     % Hyperlinks
\usepackage{graphicx}     % To include graphics
\usepackage{booktabs}     % For enhanced tables

\title{The New Age of Collusion? An Empirical Study into Airbnb's Pricing Dynamics and Market Behavior}
\author{[Dr. Richeng Piao]}
\date{\today}

\begin{document}

\maketitle
\begin{abstract}
This study investigates the implications of algorithmic pricing in digital marketplaces, focusing on Airbnb's pricing dynamics. With the advent of Airbnb's new pricing tool, this research explores how digital tools influence hosts' pricing strategies, potentially leading to market dynamics that straddle the line between efficiency and collusion. Utilizing a Regression Discontinuity Design (RDD) and Propensity Score Matching (PSM), the study examines the causal effects of the pricing tool on pricing behavior among hosts with different operational strategies. The findings aim to provide insights into the evolving landscape of digital economies, examining the balance between competitive market practices and the risk of tacit collusion facilitated by algorithmic pricing. This study contributes to the discourse on digital market regulation, offering a nuanced understanding of the implications of AI-driven tools in market dynamics and antitrust analysis.
\end{abstract}

\section{Introduction}
\subsection{Background}

In the evolving landscape of the digital economy, the emergence of sophisticated digital tools has revolutionized the way businesses operate and compete. A central development in this transformation is the advent of algorithmic pricing, which has the potential to alter market dynamics profoundly. Historically, pricing strategies have been the domain of human decision-making, often based on a mix of cost-plus logic, market sensing, and competitive analysis. However, with the rise of big data and advanced computational techniques, pricing algorithms are increasingly taking center stage.

The automation of pricing and other complex listing maintenance tasks through digital tools offers significant advantages in terms of efficiency and responsiveness to market changes. Yet, this evolution also brings forth new challenges and risks. One such risk is the potential for algorithmic collusion, where pricing algorithms could learn to implicitly coordinate with each other, leading to price settings that may not align with competitive market outcomes.

The concept of algorithmic collusion is not merely theoretical but grounded in economic principles. The Folk Theorem of game theory, for instance, suggests the possibility of multiple equilibria in infinitely repeated games, including both competitive and collusive outcomes. In digital marketplaces, where algorithms can quickly adjust prices in response to market signals, a seemingly minor event, such as a change in a single seller’s marginal cost(unilateral), could trigger a chain reaction. This reaction could potentially escalate market prices, possibly even approaching monopoly pricing levels under certain conditions.

Moreover, the self-learning capability of these algorithms, fueled by vast troves of data and machine learning techniques, could lead them to discover and sustain collusive equilibria without explicit human direction or communication. This possibility marks a departure from traditional collusion models, which typically require explicit agreements or communication between firms. In a digital setting, algorithms might independently converge on similar pricing strategies that maximize profits collectively, rather than competing aggressively as traditional market models would predict.

The potential for such outcomes is not just an academic concern but is increasingly gaining attention in policy and regulatory circles. As algorithms become more sophisticated and pervasive, understanding their impact on market dynamics, competition, and consumer welfare becomes paramount. This is especially pertinent in platforms like Airbnb, where pricing tools can provide hosts with detailed market insights, potentially influencing their pricing and listing strategies in ways that have yet to be fully understood.

Our research seeks to empirically investigate these dynamics within the context of Airbnb's marketplace. By analyzing the impact of Airbnb's new pricing tool on hosts' pricing and listing strategies, we aim to shed light on the broader implications of digital tools on market competition and potential pathways towards algorithmic collusion. This investigation is not only critical for understanding the current state of digital marketplaces but also for anticipating future developments as these technologies continue to evolve.

\subsection{Research Problem}
This research delves into the evolving interplay between technology, market dynamics, and antitrust concerns, specifically within the context of digital platforms like Airbnb. The introduction of Airbnb's new pricing tool, applicable to a broader category of digital marketplaces and sellers, serves as an ideal natural experiment for this study.

A primary focus of this research is the antitrust and anti-competitive aspects associated with AI algorithm pricing, not only in Airbnb's context but in digital marketplaces at large. The current use of similar pricing tools by humans already poses questions about competitive outcomes. However, the move towards fully automated systems, where algorithmic pricing might be automatically set, intensifies these concerns. This automation significantly reduces the cost and effort required for hosts (or general sellers in digital marketplaces) to manage their listings, potentially leading to a strategic shift from long-term, stable pricing strategies to short-term, dynamic ones.

In traditional settings, some sellers might opt for lower prices and longer-term stability due to the high costs associated with frequent price adjustments and listing management. However, the ease and efficiency offered by automated tools could encourage a more frequent adjustment of prices in response to real-time market data. This ease of pricing adjustment could lead to a strategic shift among sellers toward more dynamic pricing strategies.

The implications of this shift are profound in terms of market competition. Markets might move from competitive outcomes to collusive ones as sellers increasingly rely on algorithmic tools for pricing. These tools, designed to optimize revenues, might inadvertently facilitate a form of tacit collusion across the platform. The concern is that algorithmic pricing, especially in an infinite game setting as interpreted in the context of Folk Theorems, could sustain cooperation (or collusion) more easily than in finite settings. According to these theorems, in infinitely repeated games, multiple equilibria—including collusive equilibria—are possible, and the automation of pricing decisions could make such outcomes more likely.

This potential for algorithmic pricing tools to facilitate collusion without explicit coordination between sellers represents a significant challenge to traditional antitrust frameworks. It necessitates a nuanced understanding of how these technologies operate and their broader implications for market structure and consumer welfare.

This study aims to explore these dynamics in depth, using Airbnb's pricing tool as a case study. By empirically examining the tool's impact on pricing and listing strategies, with a particular focus on minimum stay requirements, this research seeks to offer insights into the ongoing debate on competition in digital marketplaces and the role of algorithmic pricing in market outcomes.

\section{Literature Review}
The rapid digitization of the economy and the advent of algorithmic pricing mechanisms have brought about significant changes in market dynamics. This literature review synthesizes insights from existing literature with a focus on how these developments impact market transparency, competition, and the potential for collusive behavior.

\subsection{Digitalization and Enhanced Market Transparency}
The digital transformation of the economy has greatly increased the transparency of market prices. Retail chains are increasingly posting their prices online, and brick-and-mortar stores are adopting electronic shelf labels, leading to a more dynamic pricing environment. This trend is exemplified by practices like Tesco’s price matching strategy from 2013 to 2018, which automatically adjusted prices at the till to match competitors (The Guardian). These developments, while enhancing market efficiency and consumer information, also create an environment conducive to collusion, as firms can more easily monitor and respond to competitors' pricing strategies (Rotemberg and Saloner, 1986; Green and Porter, 1984).

\subsection{Algorithmic Pricing and Market Dynamics}
The shift to algorithmic pricing, where prices are set through automated tools using data analytics and machine learning, has profound implications for market dynamics. These tools can potentially lead to more efficient market responses but also raise concerns about facilitating tacit collusion. Algorithmic pricing can enable firms to rapidly adjust to competitors' prices and market conditions, potentially leading to anti-competitive outcomes (Salcedo, 2015; Calvano et al., 2020; Klein, 2021; Hansen et al., 2021).

\subsection{Empirical Evidence on Algorithmic Collusion}
Empirical studies have begun to explore the real-world implications of algorithmic pricing and collusion. For instance, the publication of firm-specific transaction prices in the Danish ready-mixed concrete market led to decreased price competition (Albaek et al., 1997). Additionally, research indicates that gasoline prices in Germany increased significantly when local duopolies adopted algorithmic pricing strategies (Assad et al., 2020). These examples highlight the potential for algorithmic pricing to reduce market competition, even in the absence of explicit collusion among firms.

\subsection{Impact of Digital Tools on Market Competition}
Digital tools extend beyond pricing algorithms to encompass various aspects of listing and sales management in marketplaces. The automation of these tasks could lead to strategic shifts from long-term, stable pricing to short-term, dynamic pricing strategies. This shift, facilitated by the reduced cost of managing listings, could encourage more frequent price adjustments, potentially disrupting traditional market equilibria and leading to collusive outcomes (Varian, H. R., 1980, “A model of sales,” American Economic Review).

\subsection{Algorithmic Pricing and Market Competition}
\begin{itemize}
    \item A growing body of literature examines the impact of algorithmic pricing on market dynamics, with a particular focus on how these tools can influence competitive behaviors among participants.
    \item One pivotal study in this domain is "Algorithmic Pricing Facilitates Tacit Collusion Evidence," which provides empirical evidence on how algorithmic pricing can lead to less aggressive competition, and in some cases, facilitate tacit collusion among market participants.
    \item This paper's findings are particularly pertinent to the current study as they offer insights into how digital tools like Airbnb's pricing tool might influence host pricing strategies, potentially leading to more uniform and stable prices across listings, which could be indicative of tacit collusion.
\end{itemize}

\subsection{Digital Tools in Marketplaces}
\begin{itemize}
    \item The role of digital tools in shaping market competition is an emerging area of interest. These tools can increase market efficiency and transparency but also raise concerns about potential anti-competitive behaviors.
    \item The insights from "Algorithmic Pricing Facilitates Tacit Collusion Evidence Full" are instrumental in understanding the dual nature of these tools, highlighting the need for careful examination of their impact in various market contexts, including platforms like Airbnb.
\end{itemize}

\subsection{Antitrust Concerns in the Digital Era}
The rise of algorithmic pricing poses new challenges for antitrust and competition policy. Traditional antitrust frameworks, based on human decision-making, may need to be adapted to account for the potential of algorithms to facilitate tacit collusion without explicit coordination. This necessitates a nuanced understanding of how these technologies operate and their broader implications for market structure and consumer welfare (Crémer, J., de Montjoye, Y. A., Schweitzer, H., 2019, “Competition Policy for the Digital Era,” Report for the European Commission).

\section{Research Questions}

This study aims to explore the nuanced implications of Airbnb's pricing tool and its impact on market dynamics, focusing particularly on the frequency of market interaction, changes in listing characteristics, and potential anti-competitive behaviors:

\begin{enumerate}
    \item \textbf{Correlation Between Market Interaction Frequency and Pricing Strategies:}
    \begin{itemize}
        \item How do Airbnb hosts' minimum stay requirements, as an indicator of market interaction frequency, correlate with their pricing strategies, particularly in response to market dynamics? This question seeks to understand the broader relationship between the frequency of hosts' interactions with the market and their approaches to pricing, especially in the context of evolving market conditions.
    \end{itemize}

    \item \textbf{Behavioral Impact of Airbnb's New Pricing Tool:}
    \begin{itemize}
        \item How does the introduction of Airbnb's dynamic pricing tool influence the overall pricing behavior of hosts across various minimum stay requirements? This question explores the general impact of the tool on hosts' pricing strategies, assessing whether its implementation encourages more competitive pricing or trends towards tacit collusion, without explicitly focusing on hosts' operational strategies or market interaction frequency.
    \end{itemize}
    
    \item \textbf{Utilization of Airbnb's Pricing Tool and Its Impact on Hosts' Pricing Decisions:}
    \begin{itemize}
        \item Considering the availability of data on price changes and the limitations faced by fully booked hosts, how does the introduction of Airbnb's pricing tool influence the pricing decisions of hosts who are able to adjust their prices? This question aims to apply a Regression Discontinuity Design (RDD) approach to assess the causal impact of the pricing tool on hosts' pricing behaviors. Special attention will be given to comparing hosts who changed their prices with those who couldn't due to full bookings, providing insight into the tool's impact under different operational constraints.
    \end{itemize}
    
	\item \textbf{New Pricing Tool and Host Size Dynamics in Airbnb:}
	\begin{itemize}
	    \item How does Airbnb's new pricing tool affect the pricing strategies of small (fewer/smaller listings) versus large (more/larger listings) hosts, especially regarding their capacity to adjust prices in the context of varying booking statuses (fully booked vs. available)?
	    \item \textbf{Differential Price Adjustment Strategies}: This study will analyze how small and large hosts differentially adjust their prices in response to the pricing tool, focusing on their relative booking occupancy levels. It will explore whether smaller hosts, due to potentially lower booking volumes, are more agile in adapting their pricing strategies compared to larger hosts who may possess less flexibility but greater market influence.
 	   \item \textbf{Influence of Booking Status on Pricing}: Assess the impact of booking status (fully booked vs. available) on hosts' ability to utilize the pricing tool effectively. This analysis will highlight how the ability to adjust prices varies between small and large hosts, providing insights into the operational dynamics influenced by host size.
  	  \item \textbf{Economic Implications of Size and Flexibility}: The research will delve into the economic implications of host size on pricing flexibility and market responsiveness. It aims to	 understand how different-sized market participants leverage technological tools in response to policy changes, aligning with Industrial Organization principles.
    \item \textbf{Methodological Approach}: Employ Regression Discontinuity Design to discern the causal impact of the pricing tool on small and large hosts. The study will compare hosts with the ability to adjust prices (due to availability) against those who cannot (being fully booked) across both host categories. This approach provides a comprehensive analysis of the policy's impact across different segments within Airbnb's market structure, factoring in both size and operational constraints.
	\end{itemize}

\item \textbf{New Pricing Tool Impact on New vs. Established Hosts in Airbnb:}
\begin{itemize}
    \item How does the introduction of Airbnb's dynamic pricing tool differentially impact the pricing strategies and market behaviors of new versus established hosts, particularly in terms of reputation dynamics and competitive equilibrium in the platform's marketplace?
    \item \textbf{Differential Impact on Host Segments}: This part of the study will investigate the varied responses of new hosts (less experienced, potentially less reputed) and established hosts (with a solid presence and customer base) to the dynamic pricing tool.
    \item \textbf{Adaptation to Pricing Algorithms}: Assess the adaptation mechanisms in pricing strategies among new and established hosts in response to the tool. The focus will be on whether the tool facilitates faster adaptation and competitive strategies for newer hosts and how it complements the market experience of established hosts.
    \item \textbf{Reputation and Game Theory Dynamics}: Explore the strategic role of reputation, framed within game theory, post-implementation of the pricing tool. This analysis will consider how reputation influences the pricing decisions of hosts and whether the tool mitigates or amplifies existing reputation disparities between new and established hosts.
    \item \textbf{Market Equilibrium and Competitive Dynamics}: Analyze the implications of these differential impacts for competitive equilibrium on Airbnb. This will involve examining whether the tool fosters a more vibrant and competitive marketplace or creates market imbalances that favor one host segment.
    \item \textbf{Methodological Approach}: Utilize Regression Discontinuity Design to determine the causal effect of the pricing tool's introduction on new and established hosts. The study will compare pre- and post-implementation pricing strategies and market performance indicators, while controlling for other influencing factors.
\end{itemize}

    \item \textbf{Changes in Listing Characteristics Post-Policy Implementation:}
    \begin{itemize}
        \item How have Airbnb hosts adjusted their listing characteristics, like minimum stay requirements, in reaction to the introduction of the new pricing tool? This inquiry seeks to understand if changes in operational strategies are a direct response to the tool's capabilities and market insights.
        \item Are these strategic adjustments indicative of a shift towards more competitive pricing behaviors, or do they potentially signal a decrease in market diversity and individuality among listings?
        \item Is there a trend towards uniformity in listing features and pricing strategies among hosts after the introduction of the tool, hinting at a potential reduction in competitive differentiation in the Airbnb market?
    \end{itemize}

    \item \textbf{Assessing Anti-Competitive Behavior and Market Dynamics:}
    \begin{itemize}
        \item What are the emerging patterns of pricing behavior among different host segments after the policy implementation, and do these suggest any anti-competitive practices such as tacit collusion?
        \item Does the integration of the pricing tool lead to behaviors like price-fixing, or do the observed pricing strategies align with competitive market norms, despite increased pricing coordination?
        \item In what ways has the enhanced transparency and facilitation of price adjustments by the tool influenced the competitive dynamics within the Airbnb marketplace?
    \end{itemize}

    \item \textbf{Implications for Antitrust Analysis and Market Regulation:}
    \begin{itemize}
        \item Given the analyzed pricing behaviors and market responses, what implications can be drawn for antitrust analysis in digital marketplaces like Airbnb, especially in the context of algorithmic pricing tools?
        \item How do the findings contribute to the discourse on the role of pricing transparency in fostering competitive integrity versus enabling potential collusion?
        \item What regulatory considerations or policy recommendations could be derived from the study, particularly concerning the diverse pricing strategies adopted by Airbnb hosts in response to the new pricing tool?
    \end{itemize}
\end{enumerate}

\section{Methodology}

This study employs a nuanced approach utilizing Regression Discontinuity Design (RDD) and Propensity Score Matching (PSM) to rigorously analyze the impact of Airbnb's newly implemented pricing tool on host pricing strategies and broader market dynamics.

\subsection{Data Collection and Description}
\subsubsection{Nature of the Data}
\begin{itemize}
    \item Data collection began quarterly starting from December 5th, 2022, encompassing 35 cities across the United States. A global dataset is also available for broader insights.
    \item Each data scraping cycle captured the continuous daily price decisions for each listing, spanning from the date of collection up to 356 days thereafter. This approach provides a detailed view into the evolution of dynamic pricing behaviors among hosts.
\end{itemize}

\subsubsection{Content of the Data}
\begin{itemize}
    \item The dataset includes variables such as listing prices, availability, minimum stay requirements, host characteristics, and other listing attributes.
    \item The data cover a period of 3 months before and 9 months after the introduction of Airbnb's pricing tool, enabling an assessment of its impact.
    \item Additional metadata includes reviews, booking rates, and seasonal demand indicators.
    \item The inclusion of information on listings that were fully booked serves to enrich the study’s control group, especially in the RDD analysis.
    \item Key policy changes from Airbnb that are relevant to the study include:
    \begin{itemize}
		\item \textit{Redesigned Pricing Tools:} The integration of various pricing tools within the calendar interface streamlines the price-setting experience for hosts. This enhancement in pricing transparency could significantly influence hosts' pricing strategies. In line with the folk theorem, which suggests the possibility of multiple equilibria in repeated games, this change may lead to a diverse range of equilibrium outcomes in the Airbnb market. Depending on the hosts' interactions and strategic choices, these outcomes could range from highly competitive pricing strategies to more cooperative, collusive approaches. The ease of adjusting prices, combined with greater market information, provides hosts with the flexibility to navigate between these equilibria based on their preferences and market conditions.
        \item \textit{Compare Similar Listings:} This feature enables hosts to benchmark their pricing against the average of similar listings in close proximity. While this could lead to more competitive pricing by providing hosts with clearer market rate insights, it may also instigate dynamic pricing behaviors akin to the Edgeworth cycle. For instance, a random increase in one host's marginal costs might lead others to adjust their prices accordingly. This could potentially trigger a sequence of price adjustments among hosts, as each responds to changes in others' pricing strategies, fluctuating between competitive and cooperative pricing over time.
		\item \textit{Swipe-to-Select and Yearly View in Calendar:} The introduction of these features, especially the yearly view in the calendar, has the potential to significantly influence hosts' long-term pricing strategies. By offering a more extensive overview of pricing trends and availability across an extended time frame, these tools could enhance hosts' ability to engage in strategic pricing. From an Industrial Organization perspective, this improved visibility and planning capability might lead to heightened price competition among hosts, as they are better equipped to anticipate market fluctuations and adjust their prices more effectively over the long term. On the other hand, it could also facilitate a form of tacit collusion where hosts, observing and aligning with broader pricing trends, might inadvertently stabilize prices at a higher level, reducing the intensity of price competition in the market.
    \end{itemize}
\end{itemize}

\subsubsection{Usefulness of the Data}
\begin{itemize}
    \item The detailed nature of the dataset facilitates a granular examination of how pricing strategies adapt in response to market fluctuations and policy implementations.
    \item The longitudinal aspect of the quarterly data collection allows for an assessment of persistent trends and shifts in pricing strategies over time.
    \item The diversity of geographic locations covered in the dataset ensures that the findings are applicable across a variety of market environments, enhancing the study's generalizability.
    \item The inclusion of fully booked listings provides a unique opportunity to establish a control group, enhancing the robustness of the causal inferences drawn from the RDD methodology.
\end{itemize}

\subsection{Regression Discontinuity Design (RDD)}
\begin{itemize}
    \item The RDD approach is utilized to establish a causal relationship between the introduction of Airbnb's pricing tool and changes in hosts' pricing strategies. This methodology hinges on the identification of a 'cut-off' point related to the policy implementation date, which is critical for distinguishing between the treatment and control groups.
    \item The 'cut-off' point for this study is identified around the early access release of the pricing tool on March 5th, with its broader public availability on May 25th. Listings active and making price adjustments around these dates are considered the treatment group, while those not available for booking due to being fully booked are designated as the control group.
    \item To enhance the precision of the analysis, Propensity Score Matching (PSM) will be applied to pair listings in the treatment group with similar listings in the control group. This matching will be based on characteristics such as listing features, host information, and historical pricing patterns. PSM allows for a more accurate comparison by ensuring that the listings in both groups are similar in all aspects except for the exposure to the new pricing tool.
    \item The focus of the RDD analysis will be on assessing both the immediate and longer-term impacts of the pricing tool within the Airbnb market. By comparing the pricing strategies of the treatment and control groups over time, the study aims to isolate the effect of the policy from other external factors.
    \item This rigorous methodological approach aims to provide clear insights into the causal effects of Airbnb's pricing tool on host pricing behavior, offering valuable perspectives on market dynamics and policy implications.
\end{itemize}

\subsection{Propensity Score Matching (PSM)}
\subsubsection{Estimating Propensity Scores}
\begin{itemize}
    \item Logistic regression will be employed to estimate the likelihood of each listing being available or fully booked at the time of the introduction of Airbnb's new pricing tool.
    \item The regression model incorporates variables such as location, listing type, host reputation, historical pricing patterns, and proximity to the policy implementation date to accurately estimate propensity scores.
\end{itemize}

\subsubsection{Matching Listings and Balancing Groups}
\begin{itemize}
    \item Using the calculated propensity scores, listings from the treatment group (those that adjusted prices post-policy implementation) will be matched with those in the control group (listings fully booked and unable to adjust prices).
    \item Balancing checks, including standardized mean differences and variance ratios, will be conducted to ensure that the treatment and control groups are statistically similar in observable characteristics, thereby reducing selection bias.
\end{itemize}

\subsubsection{Addressing Unmatched Listings}
\begin{itemize}
    \item Listings that cannot be adequately matched will either be excluded to maintain the integrity of the analysis or subject to separate examination to identify unique attributes or outliers.
    \item Alternative methods, such as nearest-neighbor matching or caliper matching, may be employed to refine the matching process and address any discrepancies in the initial matching.
\end{itemize}

\subsubsection{Sensitivity Analysis}
\begin{itemize}
    \item Sensitivity analyses will be conducted to evaluate the impact of varying matching parameters and criteria on the study's findings.
    \item These analyses will help in assessing the stability of the results and the potential influence of unobserved confounding variables on the causal inference.
\end{itemize}

\section{Expected Outcomes and Implications}
This study aims to provide a comprehensive understanding of the impacts of Airbnb's new pricing tool on market competition:

\begin{itemize}
    \item \textbf{Algorithmic Pricing and Market Dynamics}: This research will explore how Airbnb's pricing tool influences hosts' pricing strategies. The focus is on assessing whether the tool leads to more coordinated pricing behaviors, potentially reducing price variability across listings and contributing to a more uniform market structure.

    \item \textbf{Evidence of Tacit Collusion}: The study will analyze if the usage of Airbnb's pricing tool mirrors patterns of tacit collusion observed in other markets with algorithmic pricing, as indicated in referenced papers. Specifically, it will investigate whether the tool fosters less competitive behavior among hosts, leading to more stable but potentially higher prices.

    \item \textbf{Comparative Analysis}: A detailed comparison will be conducted between hosts who adopt the pricing tool and those who do not. This analysis will shed light on the differential impacts of algorithmic pricing tools on hosts' pricing decisions and market dynamics, offering insights into the extent of competitive versus coordinated behavior.

    \item \textbf{Policy Implications}: Findings from this study will be critical for informing competition policy in digital marketplaces. The results will delineate the balance between the efficiency gains offered by algorithmic pricing tools and their potential to foster anti-competitive practices. This will contribute to the broader discourse on regulating and monitoring digital market platforms to ensure fair and competitive practices.
\end{itemize}

The outcomes of this research are expected to enhance the understanding of the role of digital tools, particularly algorithmic pricing mechanisms, in shaping competitive dynamics in online marketplaces. It will provide valuable insights into the balance between market efficiency and the risk of collusion, with significant implications for market regulation and antitrust policy.

\section{Potential Limitations and Future Research Directions}
\subsection{Current Study Limitations}
This study, while utilizing rigorous methodologies and a comprehensive dataset, acknowledges several inherent limitations:

\begin{itemize}
    \item \textbf{Data Limitations}: The quarterly scraped data provides a detailed snapshot of market dynamics, but it may not capture extremely rapid changes or intra-quarter events that significantly influence pricing strategies. This could limit the ability to detect more subtle or immediate responses to market shifts.
    
    \item \textbf{Scope of the Study}: The primary focus is on assessing the immediate impacts of Airbnb's pricing tool on host pricing behaviors. This approach may overlook longer-term market adjustments, changes in consumer behavior, and broader economic implications that unfold over extended periods.
    
    \item \textbf{Methodological Constraints}: The RDD and PSM methods are robust statistical approaches; however, they have inherent limitations in fully isolating the causal impact of the pricing tool, particularly regarding potential unobserved confounding factors. This could affect the precision of the causal inferences drawn from the analysis.
\end{itemize}

\subsection{Directions for Future Research}
Given these limitations, future research could explore several avenues to build upon the findings of this study:

\begin{itemize}
    \item \textbf{Incorporating High-Frequency Data}: Future studies could utilize higher-frequency data, such as daily or weekly scrapings, to capture more granular market dynamics and immediate responses to policy changes or market events.
    
    \item \textbf{Long-Term Impact Analysis}: Further research could extend the time horizon to examine the long-term effects of algorithmic pricing tools on market structures, host behaviors, and consumer experiences.
    
    \item \textbf{Expanding Methodological Approaches}: Employing additional econometric techniques, such as structural modeling or time-series analysis, could provide deeper insights and help address some of the limitations associated with RDD and PSM.
    
    \item \textbf{Broader Market Studies}: Expanding the scope to include different platforms or geographic markets could offer comparative insights and enhance understanding of how digital tools impact market dynamics across various contexts.
\end{itemize}

These future research directions can significantly contribute to the evolving discourse on digital marketplaces, algorithmic pricing, and the balance between market efficiency and competitive integrity.

\subsection{Next Research Paper: Demand Estimation and Market Dynamics Using the BLP Model}
Building upon the findings of this study, future research could significantly benefit from employing the Berry, Levinsohn, and Pakes (BLP) demand estimation model in the context of Airbnb's market. This approach would enable a more nuanced analysis of both demand and supply dynamics:

\begin{itemize}
    \item \textbf{Demand and Supply Curve Estimation}: Utilize the BLP model to estimate demand and supply curves in Airbnb's market. This involves analyzing listing characteristics, booking prices, and quantities while accounting for the heterogeneity in consumer preferences and host strategies. The model can provide detailed insights into how different factors, such as price, location, and listing features, influence consumer choices and host pricing decisions.

    \item \textbf{Counterfactual Market Scenarios}: Conduct counterfactual simulations to understand the impact of policy changes on market dynamics. For instance, simulate scenarios where all listings are mandated to offer instant booking or have a minimum stay requirement of one day. Analyze how these policy shifts would affect prices, quantities, and overall market equilibrium. Such simulations can reveal the potential consequences of different regulatory or platform policy changes.

    \item \textbf{Comprehensive Market Analysis}: Expand the scope to include a comprehensive analysis of market competitiveness and efficiency. Investigate how changes in listing policies, like adjustments in minimum stay requirements or the introduction of new booking features, impact the overall market structure and competitiveness.

    \item \textbf{Exploring Market Segmentation}: Examine market segmentation by categorizing listings based on various attributes such as price range, location, and type. Analyze how different segments respond to market changes and policy interventions, providing a more granular understanding of the market.

    \item \textbf{Policy Implications and Market Design}: Utilize the insights gained from these analyses to offer recommendations for platform policies and market design. The findings could inform strategies to enhance market efficiency, foster healthy competition, and improve consumer and host welfare.

    \item \textbf{Extended Time Horizon Analysis}: Consider extending the time horizon for data collection and analysis to capture long-term market trends and responses to policy changes, enriching the understanding of the market's evolution over time.
\end{itemize}

This future research direction, encompassing demand estimation, counterfactual simulations, and comprehensive market analysis using the BLP model, will complement and extend the current study's findings. It aims to provide a holistic understanding of the complex dynamics in digital marketplaces like Airbnb and offer valuable insights for platform design and regulatory considerations.

\nocite{*}
\bibliographystyle{plain}
\bibliography{references}

\end{document}